\documentclass[letterpaper,twocolumn,prb,aps,showpacs,floatfix,superscriptaddress,nofootinbib]{revtex4-1}
\usepackage[latin1]{inputenc}
\usepackage{bm}
\usepackage{multirow}
\usepackage{amssymb}
\usepackage{amsbsy}
\usepackage{amsmath}
\usepackage{graphicx}
\usepackage{epsfig}
\usepackage{placeins}
\usepackage{ulem}
\usepackage[usenames,dvipsnames]{color}
\usepackage[colorlinks,linkcolor=blue,citecolor=blue,urlcolor=blue]{hyperref}

\makeatletter

\begin{document} 
\title{Interaction-Induced Weakening of Localization \\ in Few-Particle Disordered Heisenberg Chains}

\author{Daniel Schmidtke}
\email{danischm@uos.de}
\affiliation{Department of Physics, University of Osnabr\"uck, D-49069 Osnabr\"uck, Germany}

\author{Robin Steinigeweg} 
\affiliation{Department of Physics, University of Osnabr\"uck, D-49069 Osnabr\"uck, Germany}

\author{Jacek Herbrych}
\affiliation{Cretan Center for Quantum Complexity and Nanotechnology, Department of Physics, University of Crete,
GR-71003 Heraklion, Greece}
\affiliation{Department of Physics and Astronomy, The University of Tennessee, Knoxville, Tennessee 37996, USA}
\affiliation{Materials Science and Technology Division, Oak Ridge National Laboratory, Oak Ridge, Tennessee 37831, USA}

\author{Jochen Gemmer} 
\affiliation{Department of Physics, University of Osnabr\"uck, D-49069 Osnabr\"uck, Germany}

\begin{abstract}
We investigate real-space localization in the few-particle regime of the XXZ spin--$1/2$ chain with a random magnetic
field. Our investigation focuses on the time evolution of the spatial variance of non-equilibrium densities, as resulting for a specific class of initial states, namely, pure product states of densely packed particles. Varying the strength of both particle-particle interactions and disorder, we numerically calculate the long-time evolution of the spatial variance $\sigma(t)$. For the two-particle case, the saturation of this variance yields an increased but finite localization length, with a parameter scaling different to known results for bosons. We find that this interaction-induced increase is the stronger the more particles are taken into account in the initial condition. We further find that our non-equilibrium dynamics are clearly inconsistent with normal diffusion and instead point to subdiffusive dynamics with $\sigma(t) \propto t^{1/4}$.
\end{abstract}

\maketitle

\section{Introduction}
Non-interacting particles in a disordered potential are Anderson-localized in one dimension (1D), for any disorder strength and temperature \cite{anderson1958,Mott1968,Kramer1993,Evers2008}. Recently, it has become clear that Anderson localization is also stable against weak particle-particle interactions \cite{Basko2006,Fleishman1980}. Moreover, numerous works suggest the existence of a many-body localized (MBL) phase beyond the weak-interaction limit and even at infinite temperature \cite{Pal2010,Oganesyan2007,Luitz2015}. This MBL phase is a new state of matter with several fascinating properties, ranging from the breakdown of eigenstate thermalization \cite{Serbyn2013,Huse2014,Huse2013,huse2015,Luitz2016,Khatami2012,Mondaini2015} to the logarithmic growth of entanglement as a function of time after an initial quantum quench \cite{Prosen2008,Bardarson2012,Vosk2013}. In particular the optical conductivity features a zero dc value \cite{Gopalakrishnan2015,Steinigeweg,berkelbach2010} and a low-frequency behavior as described by Mott's law \cite{Gopalakrishnan2015}. On the experimental side, an MBL phase was recently observed for interacting fermions in a quasi-random (bichromatic) optical lattice \cite{Schreiber2015,kondov2015} as also delocalization by coupling identical 1D MBL systems \cite{Bordia2016}. Investigations of amorphous iridium-oxide indicate that MBL might play an important role for its insulating states \cite{Ovadia2015}. 

The existence of a MBL phase at finite disorder and interaction strength implies that the decrease of disorder induces a transition from a localized phase (non-ergodic, non-thermal) to a delocalized phase (ergodic, thermal) \cite{Kjall2014}. This disorder-induced transition has been under active scrutiny and different probes have been suggested \cite{Filippone} such as subdiffusive power laws in the vicinity of the critical disorder \cite{Gopalakrishnan2015} (which may or may not exist \cite{Varma,Karahalios2009,Znidaric2016,Steinigeweg,Khait2016,Barisic2012,Agarwal2015,BarLev2015}). So far, a full understanding of the MBL transition is still lacking. This lack of knowledge is also related to restrictions of state-of-the-art numerical methods. On the one hand, full exact diagonalization is restricted to small systems with only a few lattice sites, where finite-size effects are strong in disorder-free cases \cite{Steinigeweg2014,Steinigeweg2015,Shawish2004}. On the other hand, much more sophisticated methods such as time-dependent density-matrix renormalization group are restricted to times scales with sufficiently low entanglement \cite{BarLev2015, Roux2008, Hauschild}.

While the overwhelming majority of works has focused on the disorder-induced transition at half filling, much less is known on the transition induced by filling at fixed disorder strength \cite{Roux2008,Mondaini2015}. Since a single particle is localized for any finite disorder, a transition from a localized phase to a delocalized phase has to occur, if the half-filling sector is delocalized. However, {\it when} and {\it how} such a transition happens exactly constitutes a non-trivial and challenging question. So far, this question has been investigated only partially, with a remarkable focus on the special case of two interacting bosons \cite{Shepelyansky1994,Frahm1999,Krimer2011,Ivanchenko2014}. Though there are a few works containing results on two interacting fermions, see e.g. Refs.~\onlinecite{Ponomarev,Frahm2016}, there is, to the best of our knowledge, no detailed investigation for this case, especially in connection with increasing the particle number beyond two. The latter is the main focus of the work at hand.

Thus, we study this question for the XXZ spin-$1/2$ Heisenberg chain. To this end, we consider a non-equilibrium scenario. First, we prepare a pure initial state of densely packed particles (also known as bound states \cite{Ganahl2012,Fukuhara2013}), where all particles ($\uparrow$-spins) are concentrated at adjacent sites and holes ($\downarrow$-spins) are located on the other sites. Then, we calculate the evolution of the particle distribution in real time and real space, using a Runge-Kutta integration of the time-dependent Schr\"{o}dinger equation \cite{Steinigeweg2014,Steinigeweg2015,Steinigeweg2016,Elsayed2013}. While the time dependence of the distribution width allows us to study the type of dynamics in general, a convergence of this width to a constant value in the long-time limit (which may or may not exist) also allows us to extract a finite localization length. To eliminate that this length is a trivial boundary effect, we choose large system sizes. The latter are feasible for different particle numbers due to our integration scheme.

The paper is organized as follows: In Sec.~\ref{model}, we introduce the investigated model, namely a Heisenberg spin--$1/2$ chain subjected to random magnetic fields. Furthermore our main observable, i.e., the time-dependent particle density distribution arising from densely-packed initial states, comparable to so-called bound states, is discussed. Also its (dynamical) broadening in terms of standard statistical variances is presented. At the end of this section, we describe shortly our numerical methods. In Sec.~\ref{scaling}, the variance is investigated regarding the scaling behavior with particle number ranging from only one particle up to four. Sec.~\ref{localization} is dedicated to a thorough investigation of the scaling behavior of the saturation value of the aforementioned variance in the two-particle case. This scaling behavior is analyzed with respect to the particle-particle interaction and disorder strength within appropriate regimes, where we interpret finite saturation values of the variances as real-space localization lengths. In Sec.~\ref{correlations}, we provide evidence that the cases of non-interacting and interacting particles can be distinguished in terms of local density correlations where we again investigate exemplary only the two particle case. Finally we end with a short summary and conclusions in Sec.~\ref{conclusion}. In Appendix \ref{statError} we provide a thorough analysis regarding statistical error estimations for mean localization lengths and Appendix \ref{app:interaction} presents a discussion of the behavior of two-particle localization lengths in the large interaction regime in comparison to results on bosons.


\section{Model and Non-Equilibrium Densities}\label{model}

We study the XXZ spin-$1/2$ chain with a random magnetic field oriented in $z$ direction. The Hamiltonian reads (with periodic boundary conditions)
\begin{equation}
H = \sum \limits^L_{i=1} \left[ J (S^x_i S^x_{i+1} + S^y_i S^y_{i+1}
+ \Delta S^z_i S^z_{i+1}) + h_iS^z_i \right] \, ,
\end{equation}
where $S_i^{x,y,z}$ are spin-$1/2$ operators at site $i$, $L$ is the number of sites, $J>0$ is the antiferromagnetic exchange coupling constant, and $\Delta$ is the exchange anisotropy. The local magnetic fields $h_i$ are random numbers drawn from a uniform distribution in the interval $h_i\in[-W,W]$. We note that, via the Jordan-Wigner transformation~\cite{Jordan1928}, this model can be mapped onto a one--dimensional model of spinless fermions with particle-particle interactions of strength $\Delta$ and a random on-site potential of strength $h_i$. We are interested in the time evolution of the density distribution
\begin{equation}
\langle n_i(t) \rangle = \frac{1}{N} \, \text{tr} [ n_i \, \rho(t) ] \, , \quad \sum_{i=1}^L \langle n_i(t) \rangle =
1\, ,
\label{eq:ocdist}
\end{equation}
where $N$ is the number of particles, ${n}_i = S^z_i+1/2$ is the occupation-number operator at site $i$, and $\rho(t)$ is the density matrix at time $t$. In this way, we can investigate the time-dependent broadening of an initial state $\rho(0)$ in real space. In the few-particle regime, i.e., $N \ll L/2$, studied in this paper, due to short--range interactions, only initial states with a sharp concentration of particles at adjacent sites are appropriate. For such ``narrow'' initial states we can expect non-trivial dynamics, while ``broad'' initial states essentially correspond to the one-particle problem. Thus, our initial states $\rho(0) = |\psi(0) \rangle \langle \psi(0)|$ read (in the Ising basis)
\begin{equation}
|\psi(0) \rangle = \! \prod \limits^{p+N-1}_{i=p} \! S^+_i \, |\! \downarrow \ldots
\downarrow \rangle = | \ldots \downarrow \underbrace{\uparrow \ldots \uparrow}_{N} \downarrow \ldots
 \rangle\, ,
\label{eq:ini}
\end{equation} 
where $S^+_i$ is the creation operator at site $i$ and $p$ is chosen to concentrate particles ($\uparrow$-spins) around $i=L/2$. These pure states of densely packed particles describe an alignment of $N$ particles directly next to each other (known as bound states for $N \sim 2$~ \cite{Ganahl2012,Fukuhara2013} and domain walls for $N \sim L/2$ ~\cite{Hauschild, steinigeweg2006}). Note that, due to periodic boundary conditions, the specific choice of $p$ is irrelevant. However, to avoid boundary effects in the following definition it is convenient to choose $p \approx L/2$.

A central quantity of our paper is the spatial variance of the above introduced particle density distribution
\begin{equation}
\sigma^2(t) = \sum \limits^L_{i=1} i^2 \, \langle n_i(t) \rangle - \left ( \sum
\limits^L_{i=1} i \, \langle n_i(t) \rangle \right)^2 \, .
\label{eq:variance}
\end{equation}
On the one hand, the time dependence of $\sigma(t)$ yields information on the type of dynamics such as power laws $t^\alpha$ ~~\cite{Laptyeva2010,Ivanchenko2014} for sub- ($\alpha < 1/2$), normal ($\alpha = 1/2$), or super-diffusion ($\alpha > 1/2$). On the other hand, we can use the long-time value $l = \lim_{t \to \infty} \sigma(t)$ as a natural definition for the localization length. Since $l$, as well as all other quantities introduced, depend on the specific disorder realization considered, we average over a sufficiently large number of disorder realizations $r$, typically $r > 1000$, to determine the mean of $l$ within negligible statistical errors, see Appendix~\ref{statError} for details. To also ensure negligibly small finite-size effects, we set $L=100$ throughout this paper. We checked that such $L$ is sufficiently large for all quantities and time scales investigated here. Thus, for the two-particle case, i.e., $N = 2$, (with Hilbert-space dimension $\text{dim} {\cal H} = 4950$), we use full exact diagonalization (ED). For larger $N$, e.g., $N = 3$ ($\text{dim} {\cal H} = 161700$), we rely on a forward iteration of the pure state $| \psi(t) \rangle$ using fourth-order Runge-Kutta with a time step $t\, J = 0.01 \ll 1$ \cite{note1}, feasible for $L=100$ due to sparse-matrix representations of operators, see Refs.~\onlinecite{Steinigeweg2014,Steinigeweg2015,Elsayed2013} for details. As demonstrated below, the results obtained from this iterative method coincide for $N=2$ with the ED results.

\section{Scaling of the Variance}\label{scaling}

\begin{figure}[t]
\includegraphics[width=1.0\columnwidth]{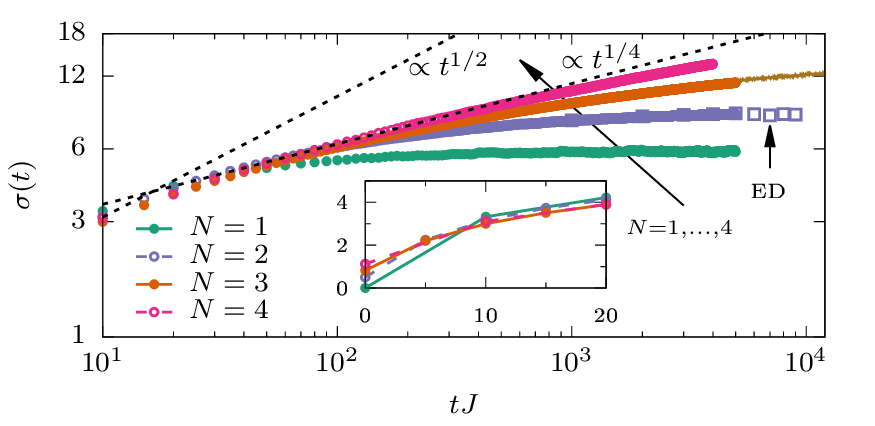}
\caption{(Color online) Log--log plot of the time evolution of the width $\sigma(t)$ for different particle numbers $N=1,\dots,4$ for $\Delta=1$, $W/J=1$. For the $N=2$ case, corresponding ED results are also shown. For the $N=3$ case, long-time data for fewer $r \approx 100$ is further depicted. Power laws are indicated for comparison. Inset: Lin--lin plot for short times.}
\label{Fig2}
\end{figure}

Now, we present our results, starting with the time evolution of the width $\sigma(t)$ and focusing on the isotropic point $\Delta = 1$ and intermediate disorder $W/J = 1$. Fig.~\ref{Fig2} summarizes $\sigma(t)$ for different particle numbers $N=1$, $2$, $3$, and $4$ in a log-log plot, with statistical errors smaller than the symbol size used (see Appendix \ref{statError}).

Several comments are in order. First, in the long-time limit, $\sigma(t)$ increases monotonously as $N$ increases from $1$ to $4$. Second, for $N=1$ and $2$, $\sigma(t)$ is approximately time-independent for $t \, J > 1000$ and takes on a constant value $\sigma < 10$, much smaller than the size of the lattice $L=100$. On the one hand, the saturation for $N=1$ is expected since in this case the actual value of $\Delta$ is irrelevant and single-particle Anderson localization persists~\cite{anderson1958,Abrahams1979}. On the other hand, the saturation for $N=2$ is in qualitative accord with corresponding results on bosons \cite{Shepelyansky1994,Frahm1999,Aizenman2009,Krimer2011,Ivanchenko2014}. Third, for $N > 2$ such conclusions are less obvious. The long time scale relevant for our dynamics, as typical for disordered problems \cite{Khatami2012,Mondaini2015,Ivanchenko2014,Luitz2016a}, systematically increases with $N$ and we do not observe a saturation of $\sigma(t)$ at the time scales depicted in Fig.~\ref{Fig2}. Note that we have also calculated $\sigma(t)$, for $N = 3$ and $r \approx 100$, up to very long $t \, J = 12000$, where it still increases significantly, see Fig.~\ref{Fig2}. This ongoing increase could be a signature of a diverging localization length and would be consistent with the delocalized phase at $N = L/2$ for this choice of $\Delta$ and $W$. Finally, the time dependence of $\sigma(t)$ is for all cases inconsistent with normal diffusion, where $\sigma(t) \propto t^{1/2}$, which has been found so far only in few disorder-free cases \cite{Steinigewg2012_3,Steinigeweg2012_2}. In fact, we find $\sigma(t) \propto t^{1/4}$ on intermediate time-intervals. The latter becomes more pronounced for larger $N$. This scaling is also expected for the chaotic spreading of nonlinear wave packets in disordered potentials \cite{Laptyeva2010,Ivanchenko2014}. Nevertheless, our data is not sufficient to state conclusively whether the observed agreement indeed indicates subdiffusion or whether it is essentially cross-over effect.

\begin{figure}[t]
\includegraphics[width=1.0\columnwidth]{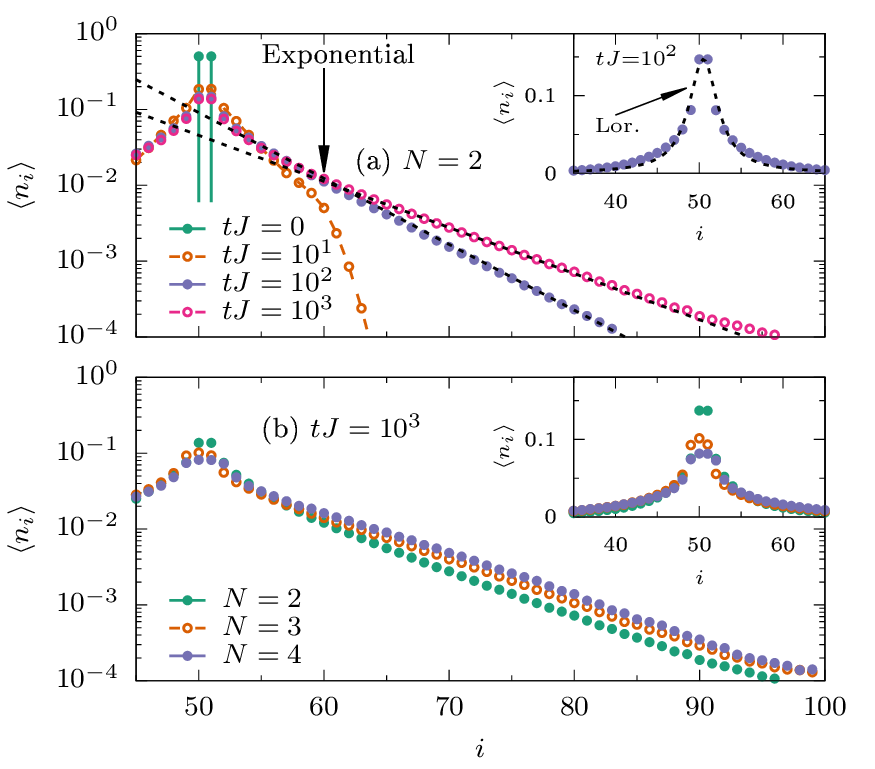}
\caption{(Color online) Site dependence of the density profile $\langle n_i(t) \rangle$ for $\Delta=1$ and $W/J=1$ at (a) times $tJ = 0,10,100$, and $1000$ for $N=2$ particles and (b) at fixed time $tJ = 1000$ for $N=2,3$ and $4$ particles, both in a semi-log plot. In (a) exponential fits to the tails are indicated. Insets: Lin-lin plots with a Lorentzian fit indicated in (a).}
\label{Fig3}
\end{figure}

In clear contrast, for short times, $\sigma(t)$ in the inset of Fig.~\ref{Fig2} is larger for smaller $N$, inverse to the long-time behavior discussed so far. This short-time behavior simply reflects that fewer particles expand in a more empty lattice. Here, disorder is not relevant yet.

To gain insight into the origin of the slow dynamics of $\sigma(t)$, we depict in Fig.~\ref{Fig3} time snapshots of the site dependence of the underlying density profile $\langle n_i(t) \rangle$ for $N=2$ in a semi-log plot. For simplicity, we focus only on the right part of the symmetric function. While the profiles are well approximated by Lorentzians around their center, see inset of Fig.~\ref{Fig3} (a), the tails show a different behavior. Remarkably, they are well described by exponentials over orders of magnitude, as expected for Anderson-localized states~\cite{anderson1958,Abrahams1979}, but here for interacting particles; see Ref.~\onlinecite{Dufour2012} for a similar decay behavior. We have found similar behavior for $N = 3$ and $4$. In fact, as shown in Fig.~\ref{Fig3} (b), the spatial decay appears to be very similar for different $N$ at fixed $t$.

\begin{figure}[t]
\includegraphics[width=1.0\columnwidth]{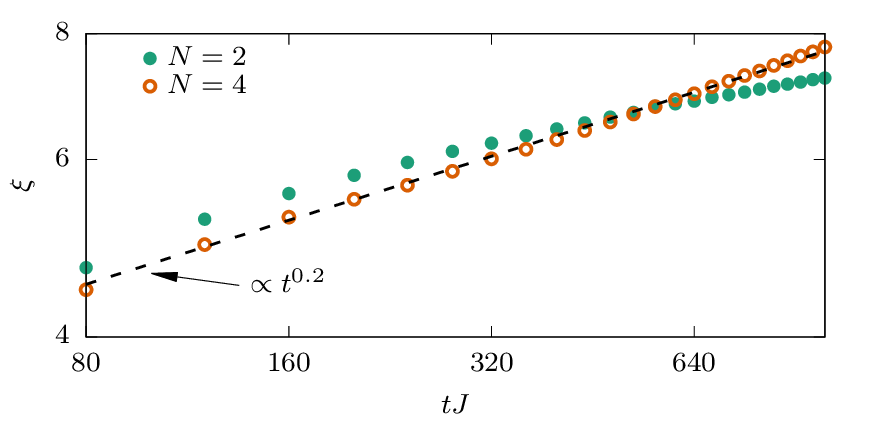}
\caption{(Color online) Log-log plot of the time-dependence of the inverse decay constant $\xi$ for $\Delta=1$ and $W/J=1$. The crossover behavior is similar to the one found for the variance $\sigma(t)$; see inset of Fig.~\ref{Fig2}. Dashed line indicates power-law fit with $\xi \propto t^{0.21}$ for $N=4$ supports sub-diffusive expansion of $\sigma(t)$ as discussed above.}
\label{Fig31}
\end{figure}

In the following, we are going to quantify this by extracting localization lengths from in Fig.~\ref{Fig3} shown exponential tails according $ \langle n_i \rangle \propto \exp(-i/\xi)$. As usual localization lengths are then denoted as the inverse decay constant, i.e., $l := \xi$. Fig.~\ref{Fig31} shows the time-evolution of $\xi$ for $N=2$ and $4$ in the time-interval $80\leq tJ \leq 1000$. Note that for $t<80$ exponential tails do not yet exist properly and density profiles are governed by the Lorentzian-type expansion at the center. Moreover since this Lorentzian-type behavior persists for all times and deviation from exponential decay may occur at the \textit{edges} of the chain, we fit the tails only for $60 \leq i \leq 90$. Remarkably, the time evolutions show a similar behavior with respect to different particle numbers as the variance $\sigma$ discussed above, i.e., at the beginning of the aforementioned time-interval the localization length is the larger the less particles are taken into account but it also increases the slower. Thus, subsequently at large times the localization length is the larger the more particles are taken into account. However, here the crossover occurs at much larger times than above. A power-law fit (see dashed lines in Fig.~\ref{Fig31}) suggests that for $N=4$ it scales as $\xi \propto t^{0.21}$ for the entire interval shown. This is rather close to a sub-diffusive expansion with $\sigma(t) \propto t^{1/4}$ and thus supports the idea that the slow growth of the tails yields sub-diffusive expansion (at least in a certain interval for the particles numbers considered here). 

\section{Localization Length}\label{localization}

\begin{figure}[b]
\includegraphics[width=1.0\columnwidth]{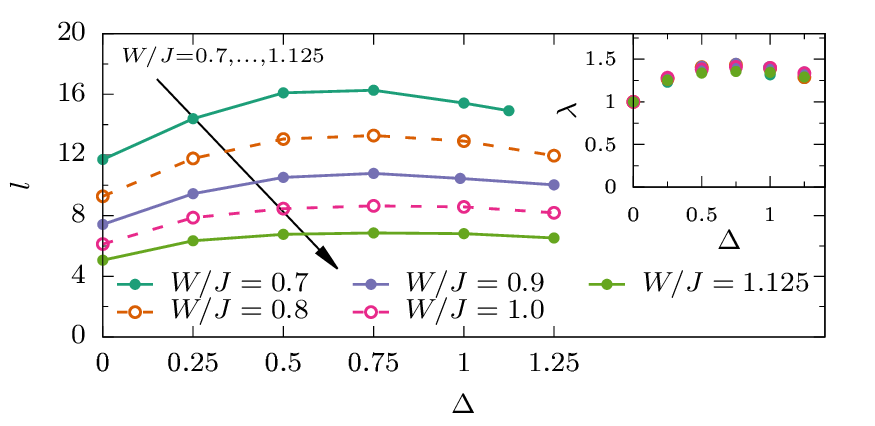}
\caption{(Color online) Dependence of the $N=2$ localization length $l$ on the interaction strength $\Delta$ for various disorders $W$, as obtained from $\sigma(t)$ at times $t\, J = 5000$. Inset: The same data as in main panel but for the relative localization length $\lambda =l(\Delta)/l(\Delta = 0)$.}
\label{Fig4}
\end{figure}

Next, we turn to the localization length $l$ which is here only analysed for $N \leq 2$ since for more particles saturation is not observed on time-scales investigated here. In Fig.~\ref{Fig4} we summarize our results for the $\Delta$ and $W$ dependence of $l$ for the two-particle case. There, we observe a clear saturation of $\sigma(t)$ at times $t \, J = 5000$ for all $\Delta$ and $W$ considered, cf. Fig.~\ref{Fig2}. Thus, $l\approx \sigma(t\, J = 5000$). In order to check saturation we also calculated $\sigma(t=10^7)$ and found $\sigma(t=10^7) \approx \sigma(5000)$; data not shown for clarity.

According to Fig.~\ref{Fig4}, the localization length $l$ is finite for all parameters depicted. Clearly, at fixed interaction strength $\Delta$, $l$ increases as disorder $W$ is decreased. At fixed $W$, $l$ increases with interaction strength for $\Delta\leq\Delta_\text{max}\approx0.75$. The decrease of $l$ for $\Delta >0.75$ occurs since our initial state is an eigenstate of the Ising limit $\Delta \to \infty$. This might be seen as running into localized states comparable to Mott states~\cite{Mott1968}. In Appendix~\ref{app:interaction}, we provide results for $W/J = 0.7,1$ for large interaction strengths, i.e., up to $\Delta \leq 14$, that visualize this behavior. Note that for $\Delta \to \infty$, bosons {\it act} like free fermions, i.e., there is no Mott insulting phase. Moreover, we find that the two-particle localization length $l_2 := l(\Delta\neq0)$ scales approximately linear with the one-particle localization length $l_1 := l(\Delta=0)$, at least in the disorder regime $0.7 \leq W \leq 1.25$. This becomes apparent from the enhancement factor $\lambda = l_2/l_1$ in Fig.~\ref{Fig4} (inset). $\lambda$ appears to be almost independently of $W$, and hence of $l_1$, and is largest in the region $\Delta \approx \Delta_\text{max}$ where $\lambda \approx 1.4$. This is significantly different from the enhancement factor for bosons with, e.g., contact interaction where $\lambda$ increases monotonously with $l_1$ at least in the regime of intermediate disorder, see e.g. Refs.~\onlinecite{Shepelyansky1994,Frahm1999,Krimer2011,Ivanchenko2014,Ponomarev,Frahm2016}. The dependence of $\lambda$ on the interaction strength for bosonic models appears to be under dispute. E.g. in Refs.~\onlinecite{Shepelyansky1994,Frahm1999,Krimer2011,Ivanchenko2014} the authors find also a monotonous increase of $\lambda$ with interaction strength, whereas in Refs.~\onlinecite{Ponomarev,Frahm2016} the authors find a similar behavior as the one at hand with a distinct interaction strength for which the enhancement is strongest (see Appendix~\ref{app:interaction}). Nevertheless in comparison to our results, there are still significant differences e.g. regarding the maximum enhancement and its distinct interaction strength; see Refs.~\onlinecite{Ponomarev,Frahm2016} for details.

\section{Local Density Correlations}\label{correlations}

Finally, we intend to shed light on the nature of the transport process and on the differences between noninteracting and interacting cases, i.e., $\Delta = 0$ and $\Delta \neq 0$, respectively. To this end, we consider the local density correlator
\begin{equation}
C_{i,\delta}(t) = \frac{\langle n_i(t) n_{i+\delta}(t) \rangle}{\langle n_i(t) \rangle
\langle n_{i+\delta}(t) \rangle}
\label{eq:corr}
\end{equation}
of site $i$ and another site in distance $\delta$, both at a given time $t$. This correlation function can be interpreted as the probability to find simultaneously site $i$ and $i+\delta$ occupied, weighted by their individual occupation probability. Thus, compared to similar correlator definitions, see e.g. Refs.~\onlinecite{Krimer2011,Fukuhara2013,Ganahl2012}, ours is apt to display correlations even if $\langle n_i(t)n_{i+\delta}(t) \rangle$ is very small; here especially in the outer tails, see Fig.~\ref{Fig3}. Note that, uncorrelated sites have $C_{i,\delta}(t) \approx 1/2$.

\begin{figure}[t]
\includegraphics[width=1.0\columnwidth]{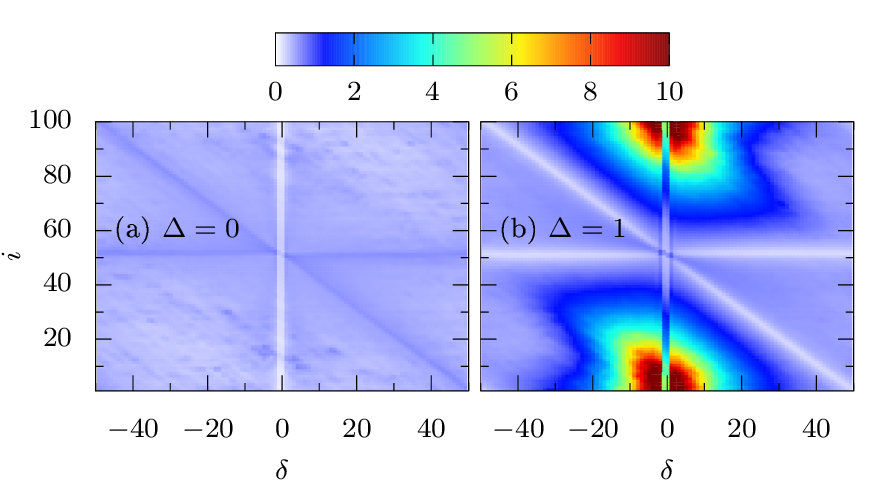}
\caption{(Color online) Local density correlator $C_{i,\delta}(t)$ vs. site $i$ and distance $\delta$ for $N=2$ particles, $W/J = 1$, and fixed time $tJ=3500$. (a) shows results for the noninteracting case $\Delta = 0$ and (b) for the interacting case $\Delta = 1$.}
\label{Fig5}
\end{figure}

In Fig.~\ref{Fig5} we show our $N=2$ results for $C_{i,\delta}(t)$ in a color map vs.\ site $i$ and distance $\delta$ for interactions turned off ($\Delta = 0$) and interaction turned on ($\Delta = 1$), focusing on long times $t \,J = 3500$ and intermediate disorder $W/J = 1$. For the $\Delta = 0$ case in Fig.~\ref{Fig5}(a), there generally is no strong enhancement of correlations. The horizontal line visible, corresponding to site $i\approx L/2$ and arbitrary $\delta$, has the slightly enhanced value $\approx 0.6$. Note that the diagonal line is equivalent to the horizontal one. Thus these lines suggest that one particle moves freely while the other remains at the initial site. For the $\Delta = 1$ case in Fig.~\ref{Fig5}(b), correlations are much more enhanced. A striking feature are strong correlations at $i \ll L/2$ and $i \gg L/2$ but with a small $\delta$. These correlations suggest that the two particles do not move independently and stay close to each other during the time evolution, in clear contrast to the noninteracting case $\Delta=0$; cf. Refs.~\onlinecite{Vlijm2015,Ganahl2012,Fukuhara2013} for results on the disorder-free case and Refs.~\onlinecite{Shepelyansky1994,Krimer2011} for bosons.

\section{Summary and Conclusions}\label{conclusion}
In summary, we investigated real-space localization in the few-particle regime of the XXZ spin-$1/2$ chain with a random magnetic field. Our investigation focused on the time evolution of the spatial variance of non-equilibrium densities, as resulting for a specific class of initial states, namely, pure states of densely packed particles. We showed that our non-equilibrium dynamics are clearly inconsistent with normal diffusion and instead point to subdiffusive dynamics. For the two-particle case, our numerical results indicated that interactions lead to an increased but still finite localization length for all parameters considered, whereas for three and four particles saturation of the variance is not observed on time-scales manageable numerically here. We also found that this interaction-induced broadening of the non-equilibrium densities is the stronger the more particles are taken into account in the initial condition. We also performed an investigation of the scaling behavior of the localization length with particle-particle interaction strength and strength of the magnetic fields where our results differ significantly from those known for bosons. Finally, we also provided evidence that the cases of non-interacting and interacting particles can be distinguished in terms of local density correlations. Our corresponding results further suggested that two interacting particles cannot move independently and stay close to each other during the time evolution in accordance with other works.

\acknowledgments
We thank F.\ Heidrich-Meisner and T.\ Prosen and for fruitful discussions. J. Herbrych acknowledges support by the U.S. Department of Energy, Office of Basic Energy Sciences, Materials Science
and Engineering Division, the EU program FP7-REGPOT-2012-2013-1 under Grant No.~316165 and thanks the University of Osnabr\"uck for kind hospitality.

\appendix

\section{Statistical Errors} \label{statError}

\begin{figure}[b]
\centering
\includegraphics[width=1.0\columnwidth]{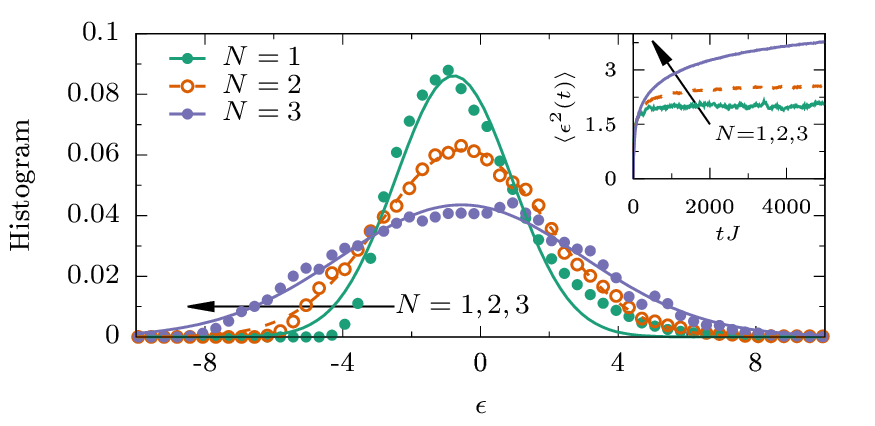}
\caption{(Color online) Histogram of $\epsilon_i(t) = \sigma_i(t) - \langle \sigma(t)\rangle$ for $N=1$, $2$, and $3$ at fixed $t \, J = 5000$ ($\Delta = 1$, $W/J = 1$, and $L=100$). Clearly, the overall shape is Gaussian (solid lines) and the width grows with $N$. Inset: Time evolution of $\langle \epsilon^2(t) \rangle$ for the same parameters.}
\label{FigS1}
\end{figure}

As the local magnetic fields $h_i$ are drawn at random, $\sigma_i(t)$ is randomly distributed around its mean $\langle
\sigma(t) \rangle$ and it is necessary to estimate statistical errors. In Fig.~\ref{FigS1} histograms for the individual deviations $\epsilon_i(t) = \sigma_i(t) - \langle \sigma (t)\rangle$ at fixed time $t \, J = 5000$, interaction $\Delta= 1$, and disorder $W/J=1$ are shown. Note that we also computed the errors for all the other parameter choices in the same way as described below. For $N = 2$ and $3$ the distributions are of Gaussian type while for $N = 1$ the distribution is slightly asymmetric. Interestingly, the distributions become broader as the particle number is increased.

In the inset of Fig.~\ref{FigS1} we show the time evolution of
\begin{equation}
\langle \epsilon^2(t) \rangle = \frac{1}{r} \sum \limits^r_{i=1} \epsilon_i(t)^2,
\end{equation}
for the same set of parameters. The time dependence of $\langle \epsilon^2(t) \rangle$ is similar to the one of $\langle\sigma(t) \rangle$ itself; however, $\langle \epsilon^2(t) \rangle < \langle \sigma(t) \rangle$. Furthermore, for $N = 1$ and $2$, the time scale where $\langle \epsilon^2(t) \rangle$ saturates at its maximum is also comparable. For $N=3$, $\langle \epsilon^2(t) \rangle$ still increases in the long-time limit, just as $\langle \sigma(t) \rangle$.

Given the Gaussian form in Fig.~\ref{FigS1}, we can estimate the error of determining $\langle \sigma(t) \rangle$ by $r$ realizations from
\begin{equation}
\sqrt{\frac{\langle \epsilon^2(t) \rangle}{r}} \, .
\end{equation}
We choose $r$ such large that this error is smaller than the symbol sizes used in the corresponding figures, i.e., typically $r > 1000$.


\begin{figure}[t]
\centering
\includegraphics[width=1.0\columnwidth]{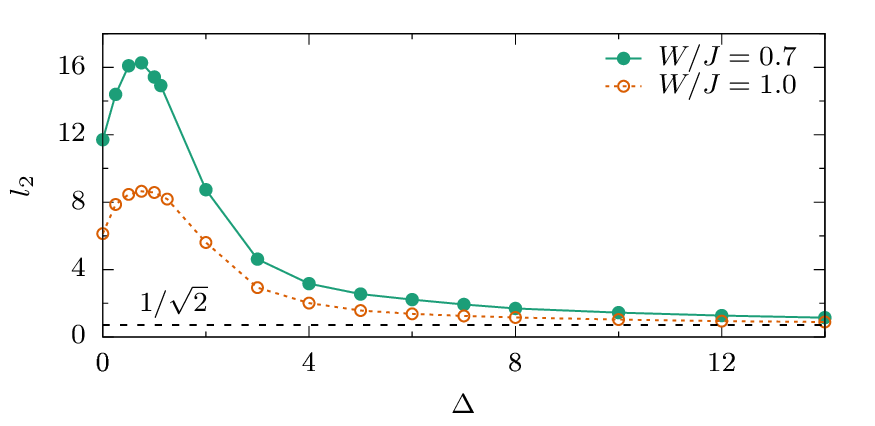}
\caption{(Color online) Two-particle localization length for $W/J = 0.7,1$ and strong interaction, i.e., $\Delta \leq 14$. For comparison the smallest localization length $l = 1/\sqrt{2}$ (which is also the initial localization of our initial states) is displayed (dashed line). Obviously, $l_2$ tends to this value in the large interaction regime, i.e., there are no significant dynamics present anymore. Statistical errors are smaller than symbol size.}
\label{FigS2}
\end{figure}
\section{Two-particle localization length for large interaction strengths} \label{app:interaction}
As pointed out in the main text, the two-particle localization length $l$ decreases if the interaction strength increases beyond a certain threshold. This fact is visualized for $W/J = 0.7,1$ and $\Delta \leq 14$ in Fig.~\ref{FigS2}. Indeed the localization lengths decreases rather strong for increasing interaction strength. For comparison, the smallest localization length, i.e., when both particles are localization on adjacent sites with $l=1/\sqrt{2}$, is displayed (dashed line). Note that our initial states are constructed to feature this value in the beginning. Clearly, for both disorder strengths the localization length tends toward this value in the large interaction regime. This also means that (detectable) dynamics of the system are not present anymore. For bosons such behavior in the high interaction regime is, as pointed out in the main text, a disputed topic. Some works propose a constant increase of the localization length with interaction strengths while others find also decreasing localization lengths as presented here. However, the differences between such results on bosons and our results remain quantitatively significant see e.g. Ref. \onlinecite{Frahm2016} for results on the same interaction strength regime. Ref. \onlinecite{Ponomarev} even shows that in the large interaction regime the enhancement vanishes, i.e., $l_2 \approx l_1$.

\bibliography{references2}

\begin{thebibliography}{10}
\bibitem{note}%
  \BibitemOpen
  \bibinfo {note} {This is a direct consequence of the indistinguishability of
  particles.}%
  \bibAnnoteFile{Stop}{note}%
\end{thebibliography}
\begin{thebibliography}{62}
\bibitem [{not()}]{note1}%
  \BibitemOpen
  \href@noop {} {}\bibinfo {note} {The build-up errors $\epsilon$ arising from
  the Runge-Kutta algorithm remain much smaller, due to the very small
  iteration step, than the symbol sizes we use. In fact $\epsilon \sim O(N
  dt^4)$ where $N$ denotes the number of iteration steps. See, e.g.,
  Ref.~\onlinecite{PhysRevB.72.020404} for details.}\BibitemShut {Stop}%
\end{thebibliography}
\end{document}